\documentclass[]{revtex4}
\usepackage{bm}
\usepackage{lettrine}
\usepackage{dcolumn}
\usepackage{bm}
\usepackage{graphicx}
\usepackage{amsmath}
\usepackage{latexsym}
\usepackage{amsfonts}
\usepackage{amssymb}
\usepackage{array}
\usepackage{epsfig}
\usepackage{epstopdf}
\usepackage{times}
\usepackage{amsmath,epsfig}
\usepackage{tikz}

\begin{document}

\title{Universal Optimal Gates Regarding Quantum Entanglement and Discord Generating}

\author{Zhe~Guan, Huan~He, Yong-Jian~Han, Chuan-Feng~Li and Guang-Can~Guo\footnote{Correspondence and requests for materials should be addressed to Y. H. (smhan@ustc.edu.cn) or C. L. (cfli@ustc.edu.cn).}}

\address{Key Laboratory of Quantum Information, Chinese Academy of Sciences,
University of Science and Technology of China, Hefei 230026, People's Republic of
China}

\begin{abstract}
{\bf
Fernando Galve \emph{et al.} $[Phys.~Rev.~Lett.~\textbf{110},~010501~(2013)]$ introduced discording power for a two-qubit unitary gate to evaluate its capability to produce quantum discord, and found that a $\pi/8$ gate has  maximal discording power. This work analyzes the entangling power of a two-qubit unitary gate, which reflects its ability to generate quantum entanglement in another way. Based on the renowned Cartan decomposition of two-qubit unitary gates, we show that the magic power of the $\pi/8$ gate produces maximal entanglement for a general value of purities for two-qubit states.
}
\end{abstract}


\maketitle

\lettrine{T}he one-qubit $\pi/8$ gate plays a special role in quantum information and quantum computation processes. As proved by Boykin \emph{et al.}~\cite{Boykin}, the $\pi/8$ gate (T, $exp(i\pi/8*\sigma_z)$ where $\sigma_z$ is the Pauli matrix), the Hadmard gate (H) and the Controlled-NOT gate (C-NOT) are universal for quantum computation. If the $\pi/8$ gate is replaced by a $\pi/4$ gate, the Gottesman-Knill theorem~\cite{N00} shows that a circuit constructed on this basis (i.e. {$T^2$, H, CNOT}) can be efficiently simulated by classical computer. The power of quantum computing using only Clifford operators can also be efficiently simulated by classical computer~\cite{gottman}, but from a one-way computer view the T-gate is a non-Clifford operator. Thus, the T-gate is the key to determine quantum computing power. Moreover, the $\pi/8$ gate plays a key role in topological quantum computation (TQC)~\cite{topological}. For general TQC systems, we can topologically protect the whole system from decoherence of the environment. However, a T-gate can not be realized by topological protected braiding and exchanging operators in many non-abelian systems such as Majorana fermions on a 2D topological superconducting medium~\cite{ludwig}. Therefore, in order to implement a universal quantum computation, we need to implement the T-gate in a topologically unprotected manner~\cite{dassarma}. As a result, any error in the unprotected T-gate will limit the threshold of the fault-tolerant quantum computation since other operators are topologically protected.

The two-qubit $\pi/8$ gate, defined as  $U_c(\pi/8,\pi/8,0)=exp(-i\pi/8(\sigma_x\otimes\sigma_x+\sigma_y\otimes\sigma_y))$, also plays a significant role in the quantum correlation of a two-qubit system. There are two main quantum correlation measurements: quantum discord and quantum entanglement. Quantum discord describes the whole quantum correlation, local and nonlocal, while entanglement only measures nonlocal quantum correlation. Other works have proposed that quantum discord, as defined by Zurek and Ollivier~\cite{discord}, acts as the source of quantum computation~\cite{AD08} and quantum state discrimination~\cite{LR11} rather than the entanglement . Recently, Fernando Galve \emph{et al.}~\cite{fer13} showed that a two-qubit $\pi/8$ gate is a perfect discorder for any value of purity $\mu$=Tr$[\rho^2]$, since such a gate produces maximally discordant mixed states (MDMS)~\cite{FG11} (as measured by the discord defined in \cite{fer13}) when acting on some special $\rho^{\mu}_{cc}$ (a classical state with zero discord). In particular, the CNOT gate corresponding to $U_c(\pi/4,0,0)$ can only have maximal discording power when $\mu\leq0.5$ and $\mu=1$, and does not generate MDMS for intermediate purity values.

Along with the ability of $U_c(\pi/8,\pi/8, 0)$ to generate maximal discord, this work shows that the $\pi/8$ gate also generates maximal quantum entanglement. Entanglement is a nonlocal resource, which has been widely investigated~\cite{Ho09} and applied to quantum information systems such as quantum communication ~\cite{Sc96}, quantum computation~\cite{Dp95,V98,N00}, and quantum teleportation~\cite{Pan01}. Based on the definition of discording power of a two-qubit unitary gate, $U$ has a similar entangling power to the system described in~\cite{fer13}. As such, we can prove that $U_c(\pi/8,\pi/8, 0)$ is also a perfect entangler that can generate maximal entanglement when acting on some special entanglement trivia states with a given purity $\mu$.

It is a bit surprising that the same gate can be both a perfect discorder and a perfect entangler. Generally, nonlocal quantum correlation is just a part of the entire quantum correlation, and is not necessarily maximized by the same gate which maximizes the whole quantum correlation, including quantum discord. The main result of this work can be written as:

\vspace{1cm}
{\bf RESULTS}

\textbf{Theorem}: $U_c(\pi/8,\pi/8,\chi), \forall \chi$ is a global two-qubit entanglement generator and has maximal entangling power for general values of purity.

More explicitly, due to Cartan decomposition and symmetry consideration, only the subset of two-qubit gates \{$U|U=e^{-i\sum_{k=x,y,z}\theta_k\sigma_k\otimes\sigma_k}$, $\theta_k \in [0, \pi/2], k=x,y,z$\} is considered. For a fixed purity $\in [1/3,1]$, $U_c(\pi/8,\pi/8, \chi)$ could find a zero-entanglement two-qubit state $\rho$, s.t., $U_c \rho U_c^\dagger$ is the maximally entangled state for this given purity.

Two key points in this theorem need emphasis:

1. Some other gates could also possibly achieve maximally entangled states for specific purities, but not all purities. In this sense, $U_c(\pi/8,\pi/8,\chi), \forall \chi$ is called "global".

2. The value of $\theta_z$ does not make a difference to our results. Normally it is set to zero.

\vspace{1cm}
{\bf DISCUSSION}

We have extended the term "entangling power" of a two-qubit unitary to another perspective, mainly focusing on its capability to generate entanglement. Our method analyzed the maximal entanglement a gate could produce when acting on zero entanglement states through the entire purity interval. We found that for certain Cartan decompositions, $U_c(\pi/8,\pi/8,0)$ is a perfect entangler which reaches the theoretical MEMS curve at all possible purities. Other gates such as the CNOT gate can only serve as perfect entanglers at low purity and unit purity. The most important result lies in the fact that $U_c(\pi/8,\pi/8,\chi), \forall\chi$ are universal optimal two-qubit unitary gates in regards to quantum entangling and discording power. This is an exciting result, and implies that a series of two-qubit unitary operators can enjoy perfect producing capability for both entanglement and discord, the two most important measures depicting quantum property.

Our work is a first attempt to quantify the power of two-qubit gates in generating entanglement in mixed states, thus providing a way to specifically analyze entanglers and disorders for general purity values. The magic power of $\pi/8$ gates to produce entanglement along with discord can provide some unique experimental utilization of $\pi/8$ gates in quantum computation and other areas.

\vspace{1cm}
{\bf METHODS}
\vspace{0.2cm}

{\bf ANALYTICAL PROOF}

To describe our main result, we need to provide a measurement for entanglement. This task is fairly important for studying quantum correlations in a system, but is not an easy thing to accomplish in a multi-particle system. Methods have been derived for quantifying entanglement in the simplest of two-qubit systems. For the entanglement of distillation, the relative entropy of entanglement and entanglement of formation (EOF), EOF is considered the canonical measure of entanglement, and is defined by calculating the von Neumann entropy of the reduced density matrix of the whole system for pure states~\cite{H97}. For mixed states, this definition is generalized as the minimum weighted sum (optimized over all possible decompositions) of the entanglement of the pure states which are the decomposition of the mixed states~\cite{C96}.

Mathematically speaking, for an arbitrary two-qubit system with density matrix $\rho$, the EOF is given by~\cite{VC00}

\begin{equation}
E(\rho)=h\left(\frac{1+\sqrt{1-C^2(\rho)}}{2}\right)
\end{equation}
Here we have the ``concurrence" $\tau=C^2(\rho)$ with $C=max\{\lambda_1-\lambda_2-\lambda_3-\lambda_4,0\}$, where $\lambda_i$ are the square roots of the eigenvalues, in decreasing order, of the matrix, $\rho(\sigma_y\otimes\sigma_y){\rho}^\ast\rho(\sigma_y\otimes\sigma_y)$, and ${\rho}^\ast$ the complex conjugation of $\rho$. Note that $h(x)$ has the form $h(x)=-x{log}_2(x)-(1-x){log}_2(1-x)$, while $E(\rho)$ is a monotonic function of $\tau$. For pure states we have $\tau=1$ and $E=1$, while $\tau=0$ corresponds to unentangled states with $E=0$.

The gates we investigate here should have a general form. We know that any two-qubit unitary operator can be expressed in Cartan form~\cite{BKJ01} by
\begin{eqnarray}
&U =(L_1\otimes L_2)U_{c}(\theta_x,\theta_y,\theta_z)(L_3\otimes L_4) \\
&U_c(\theta_x,\theta_y,\theta_z) =exp(-i\sum_{k=x,y,z}\theta_k\sigma_k\otimes\sigma_k)
\end{eqnarray}
Here, $L_i$ is local unitary which does not affect quantum entanglement, so we need only consider the Cartan kernel gate $U_c$. For the existence of symmetry, we have only considered the case of $0\leq\theta_z\leq\theta_y\leq\theta_x\leq\pi/4$.

We can define "entangling power" $\textrm{{EP}}_{\mu}(U)$ as the maximum EOF produced when acting on states with zero entanglement at purity $\mu$

\begin{equation}
\textrm{{EP}}_{\mu}(U)\equiv {max}_{\rho^{\mu}_{E=0}}E[U\rho^{\mu}_{E=0}U^{\dagger}]
\end{equation}
where $\rho^{\mu}_{E=0}$ corresponds to a two-qubit state with zero entanglement at a fixed purity $\mu$ (The term "entangling power" has been proposed in Ref.~\cite{JZ}, which  is defined as the average entanglement that the unitary operator can produce when acting on separable states). A gate $U$ can be considered as a "perfect entangler" for a given purity $\mu$ if it can produce maximally entangled mixed states (MEMS)~\cite{M01} when acting on some $\rho^{\mu}_{E=0}$. As mentioned previously, in this work we only investigate $\textrm{{EP}}_{\mu}(U_c)$.

We consider two kinds of $\rho^{\mu}_{E=0}$: the classical-classical two-qubit state $\rho^{\mu}_{cc}$ with $|\alpha_i\rangle$, $|\beta_j\rangle$ as the measurement basis elements for the system, and $p_{i,j}$ as the probability distributions at purity $\mu$ and direct product states $\rho^{\mu}_{pro}$ with A and B denoting the two qubits
\begin{eqnarray}
&& \rho^{\mu}_{cc}=\sum_{r,s}p_{i,j}|\alpha_i\rangle\langle\alpha_i|\otimes|\beta_j\rangle\langle\beta_j|, while \sum_{i,j}p^2_{i,j}=\mu\\
&& \rho^{\mu}_{pro}=\rho_{A}\otimes\rho_{B}
\end{eqnarray}
In fact, $\rho^{\mu}_{pro}$ belongs to $\rho^{\mu}_{cc}$, while $\rho^{\mu}_{cc}$ may not be within the set of $\rho^{\mu}_{pro}$ values when we consider entanglement.

To begin our proof, we first need to recall the expressions for MEMS and MDMS. According to Ref.~\cite{M01}, we can write the MEMS state $\rho_{\mathrm{ME}}(\gamma, \varphi)$ as

\begin{equation}
\rho_{\mathrm{ME}}(\gamma, \varphi)=\left(
  \begin{array}{cccc}
    1-2g(\gamma) & 0 & 0 & 0 \\
    0 & g(\gamma) & {\gamma}e^{-i\varphi} & 0 \\
    0 & {\gamma}e^{i\varphi} & g(\gamma) & 0 \\
    0 & 0 & 0 & 0\\
  \end{array}
\right)
\end{equation}
where
\begin{equation}
g(\gamma)=\left\{
  \begin{array}{ll}
    \gamma/2, & \hbox{$\gamma\geq2/3$} \\
    1/3, & \hbox{$\gamma<2/3$}
  \end{array}
\right.
\end{equation}
in Ref.~\cite{FG11} the MDMS occur when $1/2\leq\mu\leq1$ is obtained with the family $\rho_{\mathrm{MD}}(a,b,\varphi)$

\begin{equation}
\rho_{\mathrm{MD}}(a,b,\varphi)=\frac{1}{2}\left(
  \begin{array}{cccc}
    1-a+b & 0 & 0 & 0 \\
    0 & a & ae^{-i\varphi} & 0 \\
    0 & ae^{i\varphi} & a & 0 \\
    0 & 0 & 0 & 1-a-b\\
  \end{array}
\right)
\end{equation}
Here, $1/2\leq a\leq1$ and $b=1-a$. Note that both the EOF and discord do not depend on the phase factor $\varphi$, which would be canceled by a local rotation.

Interestingly, we should realize that $\rho_{\mathrm{ME}}(\gamma,\varphi)$ is just $\rho_{\mathrm{MD}}(a,b,\varphi)$ when $2/3\leq\gamma\leq1$, which means that for values of $0.556\leq\mu\leq1$, MEMS and MDMS have the same form. Conveniently, we can say that $U_c(\pi/4, 0, 0)$ can not obtain MEMS/MDMS when acting on classical-classical states according to the conclusion in Ref.~\cite{fer13}, and $U_c(\pi/8,\pi/8,0)$ is the perfect entangler/discorder since
\begin{eqnarray}
&& U_c(\pi/8,\pi/8,0)\rho_{diag}U_c^{\dagger}(\pi/8,\pi/8,0)=\rho_{\mathrm{MD}}(a,b,\pi/2)=\rho_{\mathrm{ME}}(a,\pi/2)\\
&& \rho_{diag}=\left(
    \begin{array}{cc}
      1-a &  \\
       & a \\
    \end{array}
  \right)\otimes\left(
                  \begin{array}{cc}
                    1 &  \\
                     & 0 \\
                  \end{array}
                \right)
\end{eqnarray}
This shows that the zero entanglement state $U_c(\pi/8,\pi/8,0)$ needs to act on is a direct product state. For values of $0\leq\gamma\leq2/3$ (i.e. $1/3\leq\mu\leq0.556$), we get
\begin{eqnarray}
U_c(\pi/8,\pi/8,0)\rho_{s}U_c^{\dagger}(\pi/8,\pi/8,0)=\rho_{\mathrm{ME}}(\gamma, \pi/2)\\
\rho_s=\left(
                           \begin{array}{cccc}
                             {1}/{3} &  &  &  \\
                              & 1/{3}-\gamma/2&  &  \\
                              &  & {1}/{3}+\gamma/{2} &  \\
                              &  &  & 0 \\
                           \end{array}
                         \right)
\end{eqnarray}
The proof above confirms that, for general purities, $U_c(\pi/8,\pi/8,0)$ reaches the MEMS condition, thus becoming a perfect entangler.

Another interesting phenomenon occurs when $0\leq\gamma\leq \sqrt3/3$ (i.e. $1/3\leq\mu\leq1/2$), where $U_c(\theta_x,\theta_y,0)$ with $\theta_x+\theta_y=\pi/4$ can also result in the MEMS condition. For example,
\begin{eqnarray}
&U_c(\pi/4,0,\chi)\rho_{c}U_c^{\dagger}(\pi/4,0,\chi)=\rho_{\mathrm{ME}}(\gamma, \pi/2)\\
&with\ \rho_c=\left(
                           \begin{array}{cccc}
                             {1}/{6} &  &  &-i/{6}  \\
                              & {1}/{3}-{\gamma}/{2} &  &  \\
                              &  & {1}/{3}+{\gamma}/{2} &  \\
                              {i}/{6}&  &  &{1}/{6} \\
                           \end{array}
                         \right)
\end{eqnarray}
We can equally say that when these $U_c$ act on some zero entanglement states, they can produce the maximal EOF at a given purity. Actually, the states these $U_c$ need to act on are not classical-classical states, nor direct product states. This is an exciting result, since Ref.~\cite{fer13} shows that for the same $[0,\ 1/2]$ purity interval, $U_c(\pi/4,0,0)$ is a perfect discorder while $U_c(\pi/6,\pi/12,0)$ and $U_c(\pi/5,\pi/20,0)$ are not.

\vspace{1cm}
{\bf NUMERICAL SIMULATIONS}

\begin{figure}[htbp]
\includegraphics[width=155mm]{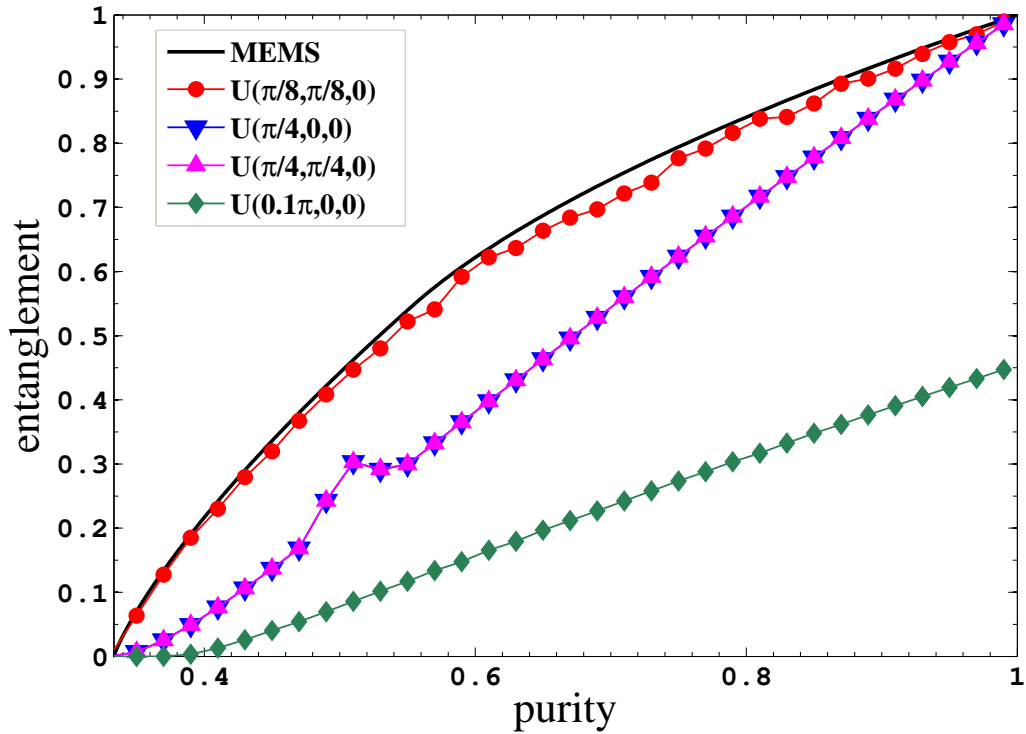}
\caption{\textbf{Entanglement power versus purity for several Cartan kernel gates based on classical-classical states.} As in Ref.~\cite{fer13}, we discrete $|\alpha_i\rangle$ and $|\beta_j\rangle$ using steps of $0.1\pi$ for each gate at any purity $\mu$, and generate $1000$ random samples for $p_{i,j}$ for a pattern of $|\alpha_i\rangle$ and $|\beta_j\rangle$. The legend, from the top, indicates: $(U_c(\pi/8,\pi/8,0)$ (red dots), $U_c(\pi/4,0,0)$ (blue inverted triangles), $U_c(\pi/4,\pi/4,0)$ (pink triangles), and $U_c(0.1\pi,0,0)$ (dark green diamonds). We see that when $\mu\in[1/3,0.556]$, the EP$(U_c(\pi/8,\pi/8,0))$ curve overlaps the theoretical curve for MEMS, which agrees with our analytical result. The loose deviation of the EP$(U_c(\pi/8,\pi/8,0))$ curve from that of MEMS when $\mu>0.556$ is due to the small possibility of obtaining product states when generating classical-classical states.}
\label{c-c}
\end{figure}

\begin{figure}[htbp]
\includegraphics[width=155mm]{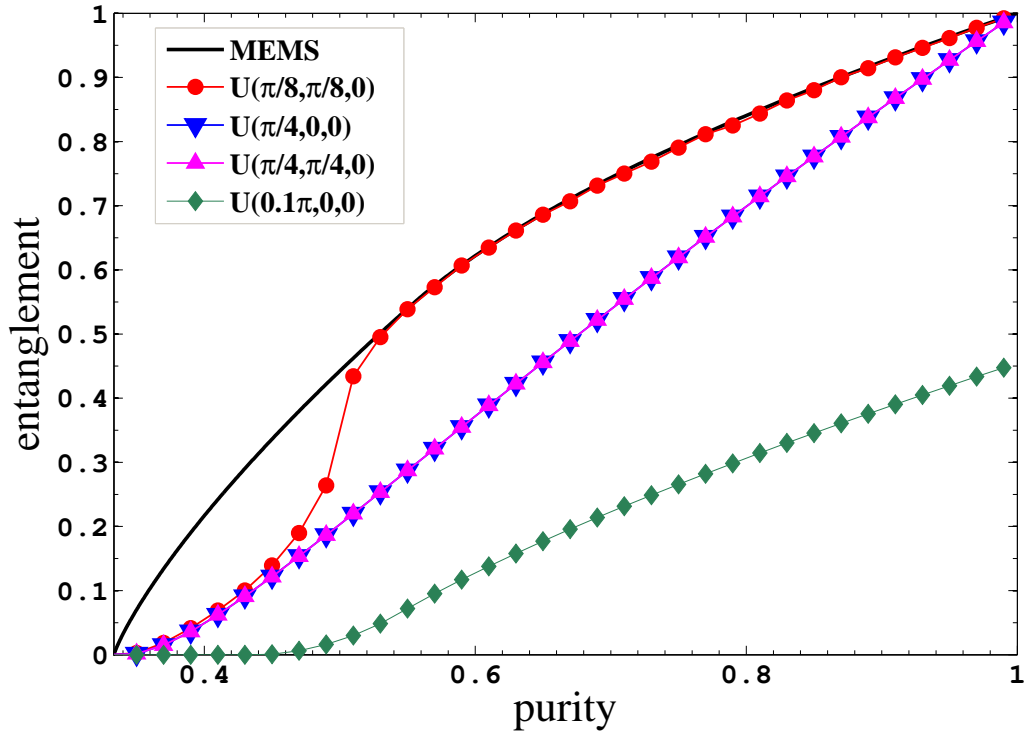}
\caption{\textbf{Entanglement power versus purity based on product states.} Recognizing the fact that there exists $\mu=\mu_{A}\times\mu_{B}$ for $\rho=\rho_{A}\otimes\rho_{B}$, we consider all possible combinations of $\mu_{A}$ and $\mu_{B}$ by steps of $0.01$ for a given $\mu$\, and generate more than $1$ million samples for each purity. When $\gamma\leq2/3$, all four gates perform poorly in producing entanglement, while the perfect overlapping of the MEMS and EP$(U_c(\pi/8,\pi/8,0))$ curves when $\mu>0.556$ serves as confirmation of our analytical result based on product states.}
\label{product}
\end{figure}

To provide some concrete examples for this discussion, we conducted numerical simulations to plot $\textrm{{EP}}_{\mu}(U_c)$ as a function of purity in the EOF-purity plane for several Cartan kernels. When a bipartite system reaches a level of mixture (here $\mu=1/3$), the entanglement disappears, a phenomenon known as entanglement sudden death (ESD)~\cite{TY04}. Therefore, the simulation was only run over the interval $1/3\leq\mu\leq1$. In developing the numerical simulation, we considered $\rho_{cc}$ and $\rho_{pro}$ separately. We first explored the performance of $\textrm{{EP}}_{\mu}(U_c)$ acting on $\rho_{cc}$. In Fig.~\ref{c-c}, we can see that $U_c(\pi/8, \pi/8, 0)$ is a perfect entangler at low purity intervals ($1/3\leq\mu\leq0.556$) since the $\textrm{{EP}}_{\mu}(U_c)$ curve overlaps with the EOF of the MEMS. However, when $\mu$ enters the $[0.556,1]$ interval, the $\textrm{{EP}}_{\mu}(U_c(\pi/8, \pi/8, 0))$ curve loosely deviates from the curve for the theoretical maximum. If we let $\textrm{{EP}}_{\mu}(U_c)$ act on $\rho_{pro}$, however, we see in Fig.~\ref{product} that $U_c(\pi/8, \pi/8, 0)$ possesses excellent entangling power when $\mu$ enters $[0.556,1]$, since the $\textrm{{EP}}_{\mu}(U_c(\pi/8, \pi/8, 0))$ curve closely matches the theoretical line. Based on these two numerical simulations, we can say that $U_c(\pi/8,\pi/8,0)$ has maximal entangling power and serves as perfect entangler at any purity $\mu$ when acting on zero entanglement states, which supports our analytical proof.

Notably, $\rho_s$ can not be written in product form as a classical-classical state. This could explain why the performance of $U_c(\pi/8,\pi/8,0)$ is poor at low purity in Fig.~\ref{product} since the simulation is based on product states. Note that $U_c(\pi/8,\pi/8,0)$ is a perfect entangler since it reaches the theoretical MEMS curve at any value of purity. When $0\leq\gamma\leq2/3$, however, $U_c(\pi/8,\pi/8,0)$ needs to act on a zero entanglement state which is not a direct product, and requires an original product state when $2/3\leq\gamma\leq1$. Thus, it is not hard to understand why the EP curve for $U_c(\pi/8,\pi/8,0)$ in Fig.~\ref{c-c} does not perfectly overlap with the MEMS curve; in a general numerical simulation of classical-classical states, the possibility of generating a direct product state is small. In our simulation based on classical-classical states and product states, $U_c(\pi/4,\pi/4,0)$ and $U_c(\pi/4,0,0)$ have the same entangling power---i.e. $U_c(\pi/4,\pi/4,0)$ is equal to $U_c(\pi/4,0,0)$(CNOT)---but their entangling power is smaller than that of $U_c(\pi/8,\pi/8,0)$. This may result from the fact that, in a one-qubit situation, a $\pi/8$ gate is more special than a $\pi/4$ gate, which can be simulated classically.

In our work, we set $\theta_z=0$ and only considered $U_c(\theta_x,\theta_y,0)$. In fact, all our results are applied to $U_c(\theta_x,\theta_y,\chi),\ \forall\chi$.

\vspace{1cm}
{\bf Acknowledgement}

This work was supported by the National Basic Research Program of China (Grant No.~ 2011CB921200), the CAS, the National Natural Science Foundation of China (Grant Nos. 11274289 and 11105135 ), the Fundamental Research Funds for the Central Universities (Nos. WK2470000011, WK2470000004, WK2470000006, and WJ2470000007). ZG and HH acknowledge support from the Fund for Fostering Talents in Basic Science of the National Natural Science Foundation of China (No.~ J1103207).

\vspace{1cm}
{\bf Author Contributions}

Z. G. and H. H. implemented specific mathematical proofs and numerical simulations. Z. G., H. H., Y-J. H. and C-F. L. wrote and modified the manuscripts. Y-J. H., C-F. L. and G-C G. supervised this project.

\vspace{1cm}
{\bf Additional Information}

\vspace{1cm}
{\bf Competing Financial Interests}

The authors declare no competing financial interests.
\end{document}